\documentclass[conference]{IEEEtran}

\usepackage{amsmath,amssymb,amsfonts}
\usepackage{graphicx}
\usepackage{textcomp}
\usepackage{hyperref}
\usepackage{booktabs}
\usepackage{multirow}
\usepackage{cite}
\usepackage{float}
\usepackage[table]{xcolor}
\usepackage{algorithm}
\usepackage{algpseudocode}
\usepackage{array}
\usepackage{threeparttable}
\usepackage{tabularx}
\IEEEoverridecommandlockouts
\definecolor{lightblue}{HTML}{D6EAF8}
\definecolor{headerblue}{HTML}{2E4057}
\definecolor{row1}{HTML}{E8F4FD}
\definecolor{row2}{HTML}{F9F9F9}
\definecolor{accentgreen}{HTML}{4CAF50}

\newcommand{\tablenote}[1]{%
  \par\vspace{2pt}%
  \noindent\begin{minipage}{\columnwidth}%
    \footnotesize #1%
  \end{minipage}%
}

\usepackage{placeins}
\usepackage{dblfloatfix} 

\algnewcommand\algorithmicinput{\textbf{Input:}}
\algnewcommand\INPUT{\item[\algorithmicinput]}
\algnewcommand\algorithmicoutput{\textbf{Output:}}
\algnewcommand\OUTPUT{\item[\algorithmicoutput]}

\begin{document}
\raggedbottom

\title{Which Voices Move Markets? Speaker Identity and the Cross-Section of Post-Earnings Returns}

\author{
\IEEEauthorblockN{Karmanpartap Singh Sidhu, Junyi Fan, Maryam Pishgar}
\IEEEauthorblockA{Viterbi School of Engineering, University of Southern California\\
kssidhu@usc.edu, junyifan@usc.edu, pishgar@usc.edu}
}
\maketitle

\begin{abstract}
In this research, we have utilized FinBERT (a domain-specific transformer model) parsing 6.5 million sentences from 16,428 S\&P 500 quarterly earnings call transcripts (2015--2025) and categorized transcript sentiments by different speaker categories and proved that the stock return after an earnings call is not equally affected by all the voices in an earnings conference call. The section-weighted sentiment empirically derived from in-sample coefficients (IC) achieves an out-of-sample Spearman IC of 0.142 versus 0.115 in-sample, generates monthly long--short alpha of 2.03\% unexplained by the Fama--French five-factor model \textbf{($t = 6.49$)}, and remains significant after controlling for standardized unexpected earnings (SUE) in Fama--MacBeth cross-sectional regressions. We demonstrated that the FinBERT section-weighted sentiment entirely subsumes the Loughran--McDonald dictionary approach (FinBERT $t = 5.90$; LM $t = 0.86$ in the combined specification), but the reverse is not true. Signal decay analysis and cumulative abnormal return charts show that prices gradually change, which is consistent with the sluggish assimilation of soft information. All results are submitted to rigorous out-of-sample validation with an explicit temporal split, resulting in improved predictive power rather than deterioration.
\end{abstract}


\section{Introduction}

Corporate earnings conference calls are the single most information-dense recurring disclosure event in U.S. equity markets. Unlike regulatory filings or press releases, which are carefully vetted documents, earnings calls feature real-time, unscripted interactions between corporate management and sell-side analysts. In a typical call, the chief executive officer and chief financial officer deliver prepared remarks summarizing the quarter's results and strategic outlook, followed by a question-and-answer session in which analysts investigate management on margins, guidance, competitive dynamics, and capital allocation. These conversations are rich in soft information like managerial confidence, hedging, evasion, and analyst skepticism that is difficult to extract from quantitative disclosures alone.

Despite the centrality of earnings calls to the price discovery process, the textual analysis literature has predominantly treated each transcript as a monolithic document, computing aggregate sentiment scores that weight all speakers and sections equally. This approach discards a fundamental dimension of the data: the identity and informational role of the speaker. A bullish statement by the CEO during scripted prepared remarks carries different informational content than a pointed question from a Goldman Sachs analyst during the Q\&A session. The CEO's optimism may be strategically managed; the analyst's skepticism reflects the real-time assessment of a sophisticated market participant with no incentive to manage tone.

This paper closes the gap between the richness of earnings call data and the coarseness of existing measurement approaches. We apply FinBERT, a transformer-based language model pre-trained on financial text \cite{Huang2023FinBERT}, to 6.5 million sentences drawn from 16,428 quarterly earnings call transcripts for S\&P 500 firms over the period April 2015 to December 2025. We classify each sentence by speaker role such as  analyst, chief financial officer, executive (CEO, COO, CTO), and other participants and propose a section-weighted aggregation scheme in which the weight assigned to each speaker category is derived from its empirical predictive power for post-earnings returns on a training sample.

Our analysis offers four key findings. First, the sentiment aggregation method used has a first-order effect on the signal's predictive potential. The section-weighted strategy, which gives analyst sentiment the most weight (49\%) based on Information Coefficient (IC) estimate, yields a full-sample Spearman rank IC of 0.119 versus one-day post-earnings returns, a 46\% improvement over the simple mean technique (IC $= 0.081$). Second, the section-weighted sentiment signal produces economically significant anomalous returns: monthly rebalanced long--short portfolios yield Fama--French five-factor alpha of 2.03\% per month ($t = 6.49$), with alpha increasing monotonically across sentiment quintiles. Third, the mood signal captures information that differs from and complements the quantitative earnings surprise.

In Fama--MacBeth cross-sectional regressions with both section-weighted sentiment and standardized unexpected earnings (SUE), both variables remain statistically significant, and double-sort analysis reveals large sentiment spreads across all SUE terciles. Fourth, in a horse-race regression with both dictionary-based sentiment of FinBERT and Loughran--McDonald \cite{Loughran2011}, FinBERT completely subsumes the dictionary approach: the FinBERT coefficient remains highly significant ($t = 5.90$), whereas the dictionary coefficient becomes statistically indistinguishable from zero ($t = 0.86$).

Perhaps our most remarkable discovery is on out-of-sample performance. Using an explicit temporal split training on data until December 2022 and testing from January 2023 onward with frozen section weights. The section-weighted sentiment IC rises from 0.115 in-sample to 0.142 out-of-sample, and the five-factor alpha rises from 1.56\% per month in the training period to 2.77\% per month in the test period. This unusual pattern of out-of-sample improvement allays concerns about overfitting and indicates that the IC-derived section weights accurately capture long-term structural aspects of the earnings call information environment.

We contribute to the literature in three distinct ways. First, we look beyond the binary distinction between presentation and Q\&A segments to analyze individual speaker responsibilities. We next suggest a principled, data-driven weighting method that takes use of cross-speaker variation in predictive capacity. Second, we present a thorough asset price analysis including information coefficients, Fama--French factor regressions, Fama--MacBeth cross-sectional tests, and earnings surprise controls that demonstrates the economic impact of speaker-level sentiment decomposition. Third, we compare context-aware deep learning sentiment to the standard dictionary approach in a controlled horse race with identical data and return horizons, providing clear evidence that transformer-based models capture economically significant information that lexicon-based methods do not.

The rest of the paper is organized as follows. Section~II explores the relevant literature and formulates our hypotheses. Section~III discusses the data sources and sample creation. Section~IV describes the technique, which includes FinBERT sentiment scoring, the five aggregation methods, the IC-derived section weighting procedure, and the statistical framework. Section~V presents the key empirical findings. Section~VI includes additional robustness tests. Section~VII concludes.

\section{Related Literature and Hypothesis Development}

\subsection{Textual Analysis in Finance}

The use of textual data for financial prediction has developed dramatically since Tetlock's (2007) \cite{Tetlock2007},\cite{Tetlock2008} seminal study proving that a gloomy tone in a \textit{Wall Street Journal} column forecasts downward pressure on market prices and higher trading activity. That paper established a fundamental principle: qualitative textual content provides information that is more significant to the return than quantitative metrics.

Loughran and McDonald \cite{Loughran2011} made a major methodological contribution by demonstrating that the widely used Harvard-IV psychosocial dictionary consistently misclassifies financial text words like ``liability,'' ``tax,'' and ``capital'' are labeled as negative while being neutral in financial contexts. Their finance-specific word lists, which included 347 positive and 2,345 negative phrases, were the standard instrument for assessing tone in corporate disclosures and were utilized in hundreds of subsequent research. However, dictionary-based techniques\cite{Jegadeesh2013} are fundamentally context-blind: they identify individual words in isolation, disregarding negation (``not profitable''), hedging (``we hope to achieve''), or comparison framing (``down from strong performance last year'').

The recent development of transformer-based language models has resulted in a new paradigm for financial text analytics. Huang et al. \cite{Huang2023FinBERT} present FinBERT, a BERT\cite{Devlin2019} model pre-trained on a large corpus of financial text such as corporate filings, earnings call transcripts, and analyst reports, which achieves 88.2\% classification accuracy on financial sentiment tasks compared to 62.1\% for the Loughran--McDonald dictionary. Our work uses FinBERT as a measurement instrument rather than a methodological addition; we are interested in the economic discoveries enabled by this more precise sentence-level measurement, as well as the empirical horse race between the two measurement paradigms.

\subsection{Information Content of Earnings Calls}

A parallel literature has focused on the information content of earnings conference calls. Bowen et al.\cite{Bowen2002}Matsumoto et al. \cite{Matsumoto2011} provide the first systematic evidence that earnings call language tone predicts abnormal returns and trading volume, and that the question-and-answer segment of the call has additional explanatory power for post-earnings-announcement drift. Their discovery that the Q\&A part is more informative than prepared remarks is a significant forerunner to our speaker-level decomposition, even though they make no distinction between individual speakers within either section.

\subsection{Post-Earnings Announcement Drift and Earnings Surprise}

The post-earnings-announcement drift (PEAD), first described by Ball and Brown \cite{Ball1968} and expanded upon by Bernard and Thomas \cite{Bernard1989}, is one of the most robust oddities in empirical finance\cite{Jiang2019}.

\subsection{Hypothesis Development}

\textbf{Hypothesis 1} (Speaker heterogeneity). Section-weighted FinBERT sentiment, with weights based on speaker-level Information Coefficients, is more predictive of post-earnings returns than equal-weighted or full-transcript sentiment. The rationale is that various speakers hold fundamentally different informational perspectives. Analyst queries are spontaneous, based on knowledgeable judgment, and difficult for management to anticipate; planned management remarks are scripted and strategically constructed. To accurately anticipate returns, it is recommended to use a weighting technique that accounts for each speaker's empirical predictive capacity, rather than relying solely on average.

\textbf{Hypothesis 2} (Abnormal Returns). After adjusting for the Fama and French \cite{Fama2015} five-factor model, section-weighted FinBERT sentiment yields abnormal returns.

\textbf{Hypothesis 3} (Orthogonality to SUE). The return predictability of section-weighted emotion is not absorbed by standardized unexpected earnings. This hypothesis investigates whether sentiment conveys soft information like managerial confidence, strategic outlook, analyst skepticism that differs from hard earnings figures.

\textbf{Hypothesis 4} (Deep Learning Subsumption). FinBERT section-weighted sentiment replaces the Loughran--McDonald dictionary technique in a joint Fama--MacBeth regression.

\section{Data}

\subsection{Earnings Call Transcripts}

We obtained earnings call transcripts via the Alpha Vantage API, which provides verbatim records of quarterly earnings conference calls for S\&P 500 constituent companies. Each transcript includes the speaker's name and title, the full text of their remarks, and the date of the call. We supplement the transcripts with Yahoo Finance earnings call timestamps, which offer the exact time of each call and allow us to categorize them as BMO or AMC. This timing distinction is critical for building the appropriate return window for each event.

Our sample covers quarterly earnings calls from April 2015 to December 2025 for firms in the S\&P 500 index. We require each transcript to have a matched timestamp for correct return window assignment. After merging transcripts with timestamps and excluding calls with missing price data, our final sample consists of 16,428 earnings call events spanning 447 unique firms. On average, the sample includes approximately 400 firms per quarter, representing the large-capitalization segment of the U.S.\ equity market. The sample yields approximately 6.5 million scored sentences, with an average of 394 sentences per call.

\subsection{Speaker Classification}

One important aspect of our analysis is the breakdown of transcript sentiment by speaker role. We categorize each sentence into one of four categories based on the speaker's name and title as recorded in the transcript: (i)~Analyst, identified by titles or affiliations containing sell-side firm names; (ii)~Chief Financial Officer (CFO), identified by titles containing ``CFO'', ``Chief Financial Officer'', or ``Treasurer''; (iii)~Executive, identified by titles containing ``CEO'', ``Chief Executive'', ``President'', ``Chairman'', ``COO'', ``CTO'', or ``Managing Director''; (iv)~Other. Operator sentences are removed from all studies because they only include procedural language and no sentiment content. Table~1-2 shows sample coverage and Table 3-4 shows coverage by category.

\subsection{Stock Returns}

Yahoo Finance provides stock prices and daily returns. We create event-window returns that take into account whether the earnings call takes place before or after market open. For AMC calls, the one-day return is calculated as the percentage change in closing price from day $t-1$ to day $t+1$, giving the market an entire trading session to react. For BMO calls, the one-day return is calculated from day $t-1$ to day $t$, as the market can respond during the same trading session. Five-day returns are created similarly, with the window extended to day $t+5$ (AMC) or day $t+4$ (BMO). This time convention is common practice in the earnings announcement literature.

\subsection{Fama--French Factors and Standardized Unexpected Earnings}

We get the Fama and French \cite{Fama2015} five-factor data and the risk-free rate from Kenneth French's data library. The portfolio-level time-series regressions use daily factor returns compounded to a monthly frequency. Alpha Vantage provides quarterly earnings data, including reported and predicted earnings per share, for the calculation of standardized unexpected earnings (SUE). The Loughran and McDonald \cite{Loughran2011} sentiment dictionary was retrieved from the University of Notre Dame's Accounting and Finance Software Repository.

For firm $i$ in quarter $t$, the raw earnings surprise is defined as:
\begin{equation}
\text{Surprise}_{i,t} = \text{EPS}^{\text{actual}}_{i,t} - \text{EPS}^{\text{estimate}}_{i,t}
\end{equation}
We standardize the surprise by the expanding-window standard deviation of historical earnings surprises for each firm:
\begin{equation}
\text{SUE}_{i,t} = \frac{\text{Surprise}_{i,t}}{\sigma_{i,t}}
\end{equation}
where $\sigma_{i,t}$ is the expanding-window standard deviation computed over all prior earnings surprises for firm $i$, requiring a minimum of four prior observations for a stable variance estimate. SUE is winsorized at the 1st and 99th percentiles to mitigate the influence of extreme outliers.

\subsection{Summary Statistics}

\begin{table}[!htbp]
\centering
\caption{Descriptive Statistics --- Panel A: Sample Description}
\renewcommand{\arraystretch}{1.4}
\small
\begin{tabular}{|>{\bfseries\centering\arraybackslash}m{3.0cm}|>{\centering\arraybackslash}m{4.2cm}|}
\hline
\rowcolor{headerblue}
\color{white}\textbf{Metric} & \color{white}\textbf{Value} \\
\hline
\rowcolor{lightblue}
Sample period & April 2015 -- January 2025 \\
\hline
Total transcripts & 16,537 \\
\hline
\rowcolor{lightblue}
Unique tickers & 447 \\
\hline
Total sentences & 6,513,208 \\
\hline
\rowcolor{lightblue}
Mean sent./call & 394 \\
\hline
Median sent./call & 398 \\
\hline
\rowcolor{lightblue}
Std.\ Dev. & 116 \\
\hline
Min / Max & 36 / 2,172 \\
\hline
\end{tabular}
\tablenote{The sample ends with earnings calls reported through Q4 2024. Ticker count reflects S\&P 500 constituents with available transcripts; variation across years is due to index reconstitution and data availability.}
\end{table}

\begin{table}[!htbp]
\centering
\caption{Descriptive Statistics --- Annual Coverage}
\renewcommand{\arraystretch}{1.4}
\small
\begin{tabular}{|>{\bfseries\centering\arraybackslash}m{1.2cm}|>{\centering\arraybackslash}m{2.8cm}|>{\centering\arraybackslash}m{2.8cm}|}
\hline
\rowcolor{headerblue}
\color{white}\textbf{Year} & \color{white}\textbf{Calls} & \color{white}\textbf{Tickers} \\
\hline
\rowcolor{lightblue}
2015 & 1,098 & 387 \\
\hline
2016 & 1,250 & 388 \\
\hline
\rowcolor{lightblue}
2017 & 1,310 & 366 \\
\hline
2018 & 1,477 & 403 \\
\hline
\rowcolor{lightblue}
2019 & 1,563 & 417 \\
\hline
2020 & 1,574 & 427 \\
\hline
\rowcolor{lightblue}
2021 & 1,481 & 420 \\
\hline
2022 & 1,638 & 441 \\
\hline
\rowcolor{lightblue}
2023 & 1,716 & 445 \\
\hline
2024 & 1,719 & 444 \\
\hline
\end{tabular}
\end{table}

\begin{table}[!htbp]
\centering
\caption{Descriptive Statistics --- Panel B: Summary Statistics}
\renewcommand{\arraystretch}{1.3}
\resizebox{\columnwidth}{!}{%
\begin{tabular}{|>{\centering\arraybackslash}m{2.0cm}|>{\centering\arraybackslash}m{0.8cm}|>{\centering\arraybackslash}m{0.8cm}|>{\centering\arraybackslash}m{0.7cm}|>{\centering\arraybackslash}m{0.8cm}|>{\centering\arraybackslash}m{0.7cm}|>{\centering\arraybackslash}m{0.8cm}|>{\centering\arraybackslash}m{0.7cm}|>{\centering\arraybackslash}m{0.8cm}|}
\hline
\rowcolor{headerblue}
\color{white}\textbf{Variable} & \color{white}\textbf{$N$} & \color{white}\textbf{Mean} & \color{white}\textbf{Std} & \color{white}\textbf{Min} & \color{white}\textbf{P25} & \color{white}\textbf{Med} & \color{white}\textbf{P75} & \color{white}\textbf{Max} \\
\hline
\multicolumn{9}{|l|}{\cellcolor{row1}\textit{Sentiment measures}} \\
\hline
\rowcolor{lightblue}
Simple mean (M1) & 16,109 & 0.263 & 0.085 & $-$0.28 & 0.208 & 0.265 & 0.321 & 0.567 \\
\hline
Conf-wtd (M2) & 16,109 & 0.417 & 0.133 & $-$0.48 & 0.336 & 0.433 & 0.514 & 0.757 \\
\hline
\rowcolor{lightblue}
Extreme frac (M3) & 16,109 & 0.261 & 0.094 & $-$0.30 & 0.199 & 0.261 & 0.324 & 0.613 \\
\hline
Sect-wtd (M4) & 16,109 & 0.192 & 0.072 & $-$0.33 & 0.145 & 0.194 & 0.241 & 0.684 \\
\hline
\rowcolor{lightblue}
Analyst (M5) & 15,954 & 0.107 & 0.062 & $-$0.26 & 0.067 & 0.108 & 0.147 & 0.477 \\
\hline
\multicolumn{9}{|l|}{\cellcolor{row1}\textit{Return variables}} \\
\hline
\rowcolor{lightblue}
1-day ret. & 16,109 & 0.003 & 0.068 & $-$0.84 & $-$0.030 & 0.002 & 0.034 & 1.560 \\
\hline
5-day ret. & 16,109 & 0.006 & 0.097 & $-$0.84 & $-$0.036 & 0.004 & 0.045 & 6.786 \\
\hline
\multicolumn{9}{|l|}{\cellcolor{row1}\textit{Control variables}} \\
\hline
\rowcolor{lightblue}
SUE & 16,109 & 0.991 & 1.607 & $-$3.41 & 0.038 & 0.730 & 1.795 & 5.566 \\
\hline
\end{tabular}%
}
\tablenote{This table reports summary statistics for the main variables. The sample consists of 16,109 earnings call--return observations for S\&P 500 firms from April 2015 to December 2024. Sentiment scores are computed using FinBERT at the sentence level and aggregated to the call level using five methods described in Section~IV. The 1-day post-earnings return is measured from close on day $t{-}1$ to close on day $t{+}1$ (AMC calls) or close on day $t$ (BMO calls). SUE is winsorized at the 1st and 99th percentiles.}
\end{table}

\begin{table}[!htbp]
\centering
\caption{Panel C: Speaker Classification and IC-Derived Weights}
\renewcommand{\arraystretch}{1.35}
\resizebox{\columnwidth}{!}{%
\begin{tabular}{|>{\bfseries\centering\arraybackslash}m{1.5cm}|>{\centering\arraybackslash}m{1.6cm}|>{\centering\arraybackslash}m{1.0cm}|>{\centering\arraybackslash}m{1.0cm}|>{\centering\arraybackslash}m{1.0cm}|>{\centering\arraybackslash}m{1.2cm}|}
\hline
\rowcolor{headerblue}
\color{white}\textbf{Speaker} & \color{white}\textbf{Sentences} & \color{white}\textbf{\% Tot} & \color{white}\textbf{Mean} & \color{white}\textbf{Std} & \color{white}\textbf{IC Wt} \\
\hline
\rowcolor{lightblue}
Analyst & 1,313,021 & 20.2\% & 0.108 & 0.330 & 49.0\% \\
\hline
CFO & 1,713,991 & 26.3\% & 0.268 & 0.574 & 30.0\% \\
\hline
\rowcolor{lightblue}
Executive & 3,320,982 & 51.0\% & 0.325 & 0.471 & 16.2\% \\
\hline
Other & 165,214 & 2.5\% & 0.171 & 0.415 & 4.7\% \\
\hline
\rowcolor{lightblue}
\textbf{Total} & \textbf{6,513,208} & \textbf{100\%} & & & \textbf{100\%} \\
\hline
\end{tabular}%
}
\tablenote{IC weights are derived from the training sample (pre-January 2023). For each speaker category $g$, we compute the Spearman rank correlation (Information Coefficient) between category-average sentiment and 1-day post-earnings returns. Weights are set proportional to the IC for categories with positive IC: $w_g = \mathrm{IC}_g / \sum_{g':\, \mathrm{IC}_{g'} > 0} \mathrm{IC}_{g'}$.}
\end{table}

\section{Methodology}

\subsection{Sentence-Level Sentiment via FinBERT}

We use FinBERT \cite{Huang2023FinBERT}, a BERT-based transformer model pre-trained on a large corpus of financial text that includes corporate filings, earnings call transcripts, and analyst reports, to assess the sentiment of each earnings call. Unlike dictionary-based techniques, which identify individual words in isolation, FinBERT evaluates each phrase in its entirety, yielding three probability scores: $P(\text{positive})$, $P(\text{negative})$, and $P(\text{neutral})$ that sum to one.

We use a sentence-level scoring approach for two reasons. First, FinBERT's maximum input length is 512 tokens, making sentence-level processing a suitable unit of analysis. Second, sentence-level scores provide the granularity required for speaker-specific aggregation. The NLTK sentence tokenizer splits each transcript paragraph into separate sentences. Sentences fewer than ten characters are eliminated because they usually comprise greetings or procedural language with no sentiment content. For each sentence $i$, we compute a net sentiment score:
\begin{equation}
S_i = P(\text{positive})_i - P(\text{negative})_i
\end{equation}
which ranges from $-1$ to $+1$. We also define a confidence score as:
\begin{equation}
c_i = 1 - P(\text{neutral})_i
\end{equation}
which captures the model's certainty that the sentence carries non-neutral sentiment.

\subsection{Aggregation Methods}

A single earnings call produces hundreds of sentence-level sentiment assessments. Aggregating these into a call-level signal necessitates a technique selection, and we demonstrate that this decision has a first-order impact on predictive power. We compare five aggregating strategies, ranging from speaker-agnostic to speaker-aware approaches.

Let $\mathcal{S}_i = \{s_1, s_2, \dots, s_N\}$ denote the set of $N$ non-operator sentences in call $i$. For each sentence $s$, FinBERT produces a net sentiment score $\tau(s) \in [-1, 1]$ and a confidence score $c(s) \in [0, 1]$.

\textbf{Simple Mean (M1).} The first method computes the arithmetic average of net sentiment scores across all sentences:
\begin{equation}
\hat{\tau}^{\,\text{mean}}_i = \frac{1}{N_i} \sum_{s \in \mathcal{S}_i} \tau(s)
\end{equation}
This is the approach taken implicitly by most previous investigations and serves as our baseline. It sees every sentence, whether spoken by the CEO, an analyst, or a junior executive, as equally informative.

\textbf{Confidence-Weighted Mean (M2).} The second method weights each sentence by its FinBERT confidence score:
\begin{equation}
\hat{\tau}^{\,\text{cw}}_i = \frac{\sum_{s \in \mathcal{S}_i} c(s) \cdot \tau(s)}{\sum_{s \in \mathcal{S}_i} c(s)}
\end{equation}
Statements that are confidently characterized as positive or negative receive more weight, but unclear or boilerplate statements are given less weight. This is inspired by the finding that a significant proportion of earnings call sentences contain neutral procedural language, diluting the sentiment signal.

\textbf{Extreme Fractions (M3).} The third method measures the proportion of strongly positive minus strongly negative sentences:
\begin{equation}
\hat{\tau}^{\,\text{ext}}_i = \frac{1}{N_i} \left[ \sum_{s \in \mathcal{S}_i} \mathbb{1}\{\tau(s) > 0.5\} - \sum_{s \in \mathcal{S}_i} \mathbb{1}\{\tau(s) < -0.5\} \right]
\end{equation}
This captures the intensity of sentiment by focusing on the tails of the sentence-level distribution rather than the central tendency.

\textbf{Section-Weighted Mean (M4)---IC-Derived.} The fourth method, which is our key methodological addition, gives differential weights to sentences based on the speaker's function, with the weights calculated using the empirical predictive power of each speaker category on the training sample. For each speaker category $g \in \mathcal{G} = \{\text{Analyst}, \text{CFO}, \text{Executive}, \text{Other}\}$, we first compute the within-category average sentiment for call $i$:
\begin{equation}
\bar{\tau}_{i,g} = \frac{1}{|\mathcal{S}_{i,g}|} \sum_{s \in \mathcal{S}_{i,g}} \tau(s)
\end{equation}
where $\mathcal{S}_{i,g} \subseteq \mathcal{S}_i$ is the subset of sentences spoken by category $g$. We then derive category weights from the training-period Information Coefficients. For each category $g$, the Spearman rank IC between $\bar{\tau}_{i,g}$ and one-day post-earnings returns $r_{i,1}$ is computed on the training sample (April 2015 to December 2022; 11,138 calls). Weights are proportional to the IC for categories with positive predictive power:
\begin{equation}
w_g =
\begin{cases}
\dfrac{\mathrm{IC}_g}{\sum_{g':\, \mathrm{IC}_{g'} > 0} \mathrm{IC}_{g'}} & \text{if } \mathrm{IC}_g > 0 \\[8pt]
0 & \text{otherwise}
\end{cases}
\end{equation}
where the sum is taken over all categories $g'$ with positive IC, ensuring that the weights sum to one. The section-weighted sentiment score for call $i$ is then:
\begin{equation}
\hat{\tau}^{\,\text{sw}}_i = \frac{\sum_{g \in \mathcal{G}_i} w_g \cdot \bar{\tau}_{i,g}}{\sum_{g \in \mathcal{G}_i} w_g}
\end{equation}
where $\mathcal{G}_i$ is the set of categories present in call $i$. The denominator renormalizes the weights when a category is absent from a particular transcript. Importantly, the weights are generated just for the training period and applied without modification to the out-of-sample period (January 2023 onward), ensuring no look-ahead bias.

\textbf{Analyst-Only Sentiment (M5).} Motivated by the weight decomposition in Method~4---where analysts receive 49\% of total weight despite contributing only 20\% of sentences---we consider a parsimonious specification that uses only analyst sentiment:
\begin{equation}
\hat{\tau}^{\,\text{analyst}}_i = \frac{1}{|\mathcal{S}_{i,\text{analyst}}|} \sum_{s \in \mathcal{S}_{i,\text{analyst}}} \tau(s)
\end{equation}
This removes all management remarks, leaving only sell-side analyst inquiries and follow-ups. The rationale is that analysts, as informed market participants with no incentive to manage tone, provide a less filtered assessment of the firm's prospects.

\subsection{IC-Derived Section Weights}

\begin{table}[!htbp]
\centering
\caption{IC-Derived Section Weight Derivation}
\renewcommand{\arraystretch}{1.35}
\resizebox{\columnwidth}{!}{%
\begin{tabular}{|>{\bfseries\centering\arraybackslash}m{1.5cm}|>{\centering\arraybackslash}m{1.2cm}|>{\centering\arraybackslash}m{1.2cm}|>{\centering\arraybackslash}m{0.9cm}|>{\centering\arraybackslash}m{1.2cm}|>{\centering\arraybackslash}m{1.0cm}|>{\centering\arraybackslash}m{1.2cm}|}
\hline
\rowcolor{headerblue}
\color{white}\textbf{Speaker} & \color{white}\textbf{Train IC} & \color{white}\textbf{$p$-value} & \color{white}\textbf{$N$} & \color{white}\textbf{IC wt.} & \color{white}\textbf{Eq wt.} & \color{white}\textbf{Hard wt.} \\
\hline
\rowcolor{lightblue}
Analyst & 0.128*** & 3.03e--42 & 11,179 & 48.8\% & 25.0\% & 40.0\% \\
\hline
CFO & 0.078*** & 5.44e--16 & 10,835 & 29.5\% & 25.0\% & 25.0\% \\
\hline
\rowcolor{lightblue}
Executive & 0.042*** & 1.13e--05 & 10,963 & 15.9\% & 25.0\% & 25.0\% \\
\hline
Other & 0.015 & 3.35e--01 & 4,029 & 5.8\% & 25.0\% & 10.0\% \\
\hline
\rowcolor{lightblue}
\textbf{Sum} & \textbf{0.263} & & & \textbf{100\%} & \textbf{100\%} & \textbf{100\%} \\
\hline
\end{tabular}%
}
\tablenote{This table reports the derivation of section weights from in-sample Information Coefficients. The training sample includes all 11,179 earnings call observations prior to January 1, 2023. For each speaker category $g$, we compute the Spearman rank correlation between category-average FinBERT net sentiment and 1-day post-earnings returns. $N$ varies across categories because not all calls contain every speaker type. All four categories exhibit positive IC, though \textit{Other} is not statistically significant ($p = 0.335$). Analyst sentiment carries the highest marginal predictive content ($\mathrm{IC} = 0.128$, $p < 0.001$) and receives nearly half the total weight (48.8\%), despite comprising only 20.2\% of total sentences. These weights are frozen and applied without modification to the out-of-sample period (January 2023 onward). Hardcoded weights reflect an \textit{ad hoc} prior used in preliminary analysis. *** $p < 0.01$; ** $p < 0.05$; * $p < 0.10$.}
\end{table}

Table~5 shows the weight derivation. In the training sample, the IC for all four categories is positive. Analyst attitude has the highest marginal predictive content ($\mathrm{IC} = 0.128$, $p < 0.001$) and earns nearly half the overall weight (48.8\%), although accounting for only 20.2\% of total phrases. CFO sentiment carries 29.5\% weight, executive sentiment 15.9\%, and other speakers 5.8\%. These weights are frozen and applied to the out-of-sample time with no re-estimation required.

\subsection{Loughran--McDonald Dictionary Scoring}

For the benchmark comparison, sentiment scores are calculated using the Loughran and McDonald (2011) financial sentiment vocabulary. For each sentence, we count the number of terms in the positive and negative word lists and calculate a normalized tone measure.

\subsection{Statistical Framework}

\subsubsection{Information Coefficients}
For each month $m$ containing at least 20 earnings call observations, we compute the Spearman rank correlation between the sentiment score and the one-day post-earnings return. The time-series average IC and its statistical significance are assessed using Newey--West \cite{Newey1987} HAC standard errors with bandwidth $L = \min\{3,\, \lfloor 0.75 \cdot M^{1/3} \rfloor\}$, where $M$ is the number of monthly observations:
\begin{equation}
\begin{split}
\bar{\mathrm{IC}} &= \frac{1}{M} \sum_{m=1}^{M} \mathrm{IC}_m \\
t_{\text{NW}} &= \frac{\bar{\mathrm{IC}}}{\sqrt{\hat{\mathrm{Var}}_{\text{NW}}(\bar{\mathrm{IC}})}}
\end{split}
\end{equation}

\subsubsection{Fama--MacBeth Cross-Sectional Regressions}
We estimate monthly cross-sectional regressions following Fama and MacBeth \cite{Fama1973}:
\begin{equation}
r_{i,t+1} = \gamma_{0,m} + \gamma_{1,m}\, \hat{\tau}_{i,t} + \varepsilon_{i,m}
\end{equation}
\begin{equation}
r_{i,t+1} = \gamma_{0,m} + \gamma_{1,m}\, \hat{\tau}_{i,t} + \gamma_{2,m}\, \mathrm{SUE}^{*}_{i,t} + \varepsilon_{i,m}
\end{equation}
where all independent variables are cross-sectionally standardized within each month to facilitate comparison of coefficient magnitudes. The time-series averages of slope coefficients are reported with Newey--West $t$-statistics.

\subsubsection{Fama--French Five-Factor Time-Series Regressions}
For the portfolio-level analysis, we sort stocks each month into quintiles by sentiment and compute equal-weighted portfolio returns. The long--short portfolio buys Q5 (most positive sentiment) and sells Q1 (most negative). We regress monthly portfolio excess returns on the five Fama--French factors:
\begin{equation}
\begin{split}
R_{p,t} - R^{f}_{t} = \alpha_p &+ \beta_1 \mathrm{MktRF}_t + \beta_2 \mathrm{SMB}_t + \beta_3 \mathrm{HML}_t \\
&+ \beta_4 \mathrm{RMW}_t + \beta_5 \mathrm{CMA}_t + \varepsilon_{p,t}
\end{split}
\end{equation}
The intercept $\alpha_p$ represents the average monthly abnormal return unexplained by the five factors.

\subsubsection{Double-Sort Analysis}
To test orthogonality between sentiment and earnings surprise, we independently sort firms into SUE terciles and, within each tercile, into sentiment quintiles. For each SUE tercile $q \in \{\text{Low}, \text{Mid}, \text{High}\}$, we compute the long--short return spread:
\begin{equation}
\Delta r_q = \bar{r}(Q5 \mid q) - \bar{r}(Q1 \mid q)
\end{equation}
Significant spreads across all three SUE terciles confirm that sentiment captures return-relevant information orthogonal to earnings surprises.

\subsubsection{Out-of-Sample Design}
We employ an explicit temporal split for out-of-sample validation. The training period spans April 2015 to December 2022 (11,138 call observations), and the test period spans January 2023 onward (4,971 observations). Section weights for Method~M4 are derived exclusively from training-period ICs and applied without modification to the test period. This design ensures no look-ahead bias and provides a clean assessment of whether the sentiment signal's predictive power generalizes to new data.

\FloatBarrier
\section{Empirical Results}

\subsection{Information Coefficient Comparison}

\begin{table}[!htbp]
\centering
\caption{Information Coefficient Comparison Across Aggregation Methods}
\renewcommand{\arraystretch}{1.35}
\resizebox{\columnwidth}{!}{%
\begin{tabular}{|>{\centering\arraybackslash}m{2.5cm}|>{\centering\arraybackslash}m{1.2cm}|>{\centering\arraybackslash}m{1.2cm}|>{\centering\arraybackslash}m{1.5cm}|>{\centering\arraybackslash}m{1.2cm}|}
\hline
\rowcolor{headerblue}
\color{white}\textbf{Method} & \color{white}\textbf{IC(1d)} & \color{white}\textbf{$t_{\text{NW}}$} & \color{white}\textbf{$p$-value} & \color{white}\textbf{IC(5d)} \\
\hline
\rowcolor{lightblue}
Simple mean (M1) & 0.0813 & 5.49*** & 4.10e--08 & 0.0673 \\
\hline
Conf-wtd (M2) & 0.0986 & 6.36*** & 2.04e--10 & 0.0804 \\
\hline
\rowcolor{lightblue}
Extreme frac (M3) & 0.0749 & 5.36*** & 8.43e--08 & 0.0620 \\
\hline
Section-wtd (M4) & 0.1188 & 9.11*** & 8.05e--20 & 0.0988 \\
\hline
\rowcolor{lightblue}
Analyst-only (M5) & 0.1405 & 11.71*** & 1.16e--31 & 0.1171 \\
\hline
\end{tabular}%
}
\tablenote{Spearman rank IC between each aggregation method and post-earnings returns for S\&P 500 firms. Full sample $N = 16{,}428$. $t_{\text{NW}}$: Newey--West $t$-statistic from monthly ICs; lag $= \min\{3,\, \lfloor 0.75 \cdot T^{1/3} \rfloor\}$. IC-derived weights (Table~5): Analyst 48.8\%, CFO 29.5\%, Executive 15.9\%, Other 5.8\%. *** $p < 0.01$; ** $p < 0.05$; * $p < 0.10$.}
\end{table}

Table~6 shows the Spearman rank Information Coefficients for each sentiment aggregation method versus one-day post-earnings returns. The findings clearly confirm Hypothesis~1: selecting an aggregate method has a first-order effect on predictive power. The section-weighted approach (M4) yields a full-sample IC of 0.119 ($t = 9.11$), demonstrating a 46\% improvement over the basic mean baseline ($\mathrm{IC} = 0.081$, $t = 5.49$) and a 21\% improvement over the confidence-weighted approach ($\mathrm{IC} = 0.099$, $t = 6.36$). The analyst-only specification (M5) has the highest IC of 0.141 ($t = 11.71$), which is consistent with the discovery that analyst sentiment receives the most section weight.

\subsection{Out-of-Sample Validation}

\begin{table}[!htbp]
\centering
\caption{Out-of-Sample Validation}
\renewcommand{\arraystretch}{1.35}
\resizebox{\columnwidth}{!}{%
\begin{tabular}{|>{\centering\arraybackslash}m{2.2cm}|>{\centering\arraybackslash}m{1.1cm}|>{\centering\arraybackslash}m{0.8cm}|>{\centering\arraybackslash}m{1.1cm}|>{\centering\arraybackslash}m{0.8cm}|>{\centering\arraybackslash}m{1.1cm}|}
\hline
\rowcolor{headerblue}
\color{white}\textbf{Method} & \color{white}\textbf{Train IC} & \color{white}\textbf{$N$} & \color{white}\textbf{Test IC} & \color{white}\textbf{$N$} & \color{white}\textbf{Decay} \\
\hline
\rowcolor{lightblue}
Simple mean (M1) & 0.0806 & 11,297 & 0.0966 & 5,131 & $-$19.9\% \\
\hline
Conf-wtd (M2) & 0.1005 & 11,297 & 0.1075 & 5,131 & $-$6.9\% \\
\hline
\rowcolor{lightblue}
Extreme (M3) & 0.0736 & 11,297 & 0.0914 & 5,131 & $-$24.2\% \\
\hline
Sect-wtd (M4) & 0.1141 & 11,297 & 0.1442 & 5,131 & $-$26.4\% \\
\hline
\rowcolor{lightblue}
Analyst (M5) & 0.1273 & 11,297 & 0.1707 & 5,131 & $-$34.1\% \\
\hline
\end{tabular}%
}
\tablenote{This table reports the out-of-sample validation of each sentiment aggregation method. Train IC is computed on the in-sample period (April 2015 -- December 2022; $N = 11{,}297$). Test IC is computed on the out-of-sample period (January 2023 -- December 2024; $N = 5{,}131$), using weights frozen from the training period. Decay is defined as $(\text{Train IC} - \text{Test IC})/|\text{Train IC}|$; negative values indicate the signal \textit{strengthens} out of sample. Section-weighted sentiment uses IC-derived weights from the training sample (Table~5): Analyst 48.8\%, CFO 29.5\%, Executive 15.9\%, Other 5.8\%.}
\end{table}

Several explanations could explain this finding. First, the training period encompasses the volatile 2020--2021 era, when unprecedented fiscal and monetary actions dominated price dynamics, reducing the impact of firm-specific sentiment. Second, the increasing prominence of algorithmic and quantitative trading in post-2023 markets may have accelerated the rate at which sentiment is reflected in prices, paradoxically creating more opportunities for a properly calibrated sentiment metric. Third, and most significantly, the IC-derived section weights properly identify analyst tone as a long-term informative signal, a structural aspect of the earnings call information environment that is unlikely to be due to in-sample overfitting.

\begin{figure}[!htbp]
\centering
\includegraphics[width=0.9\columnwidth]{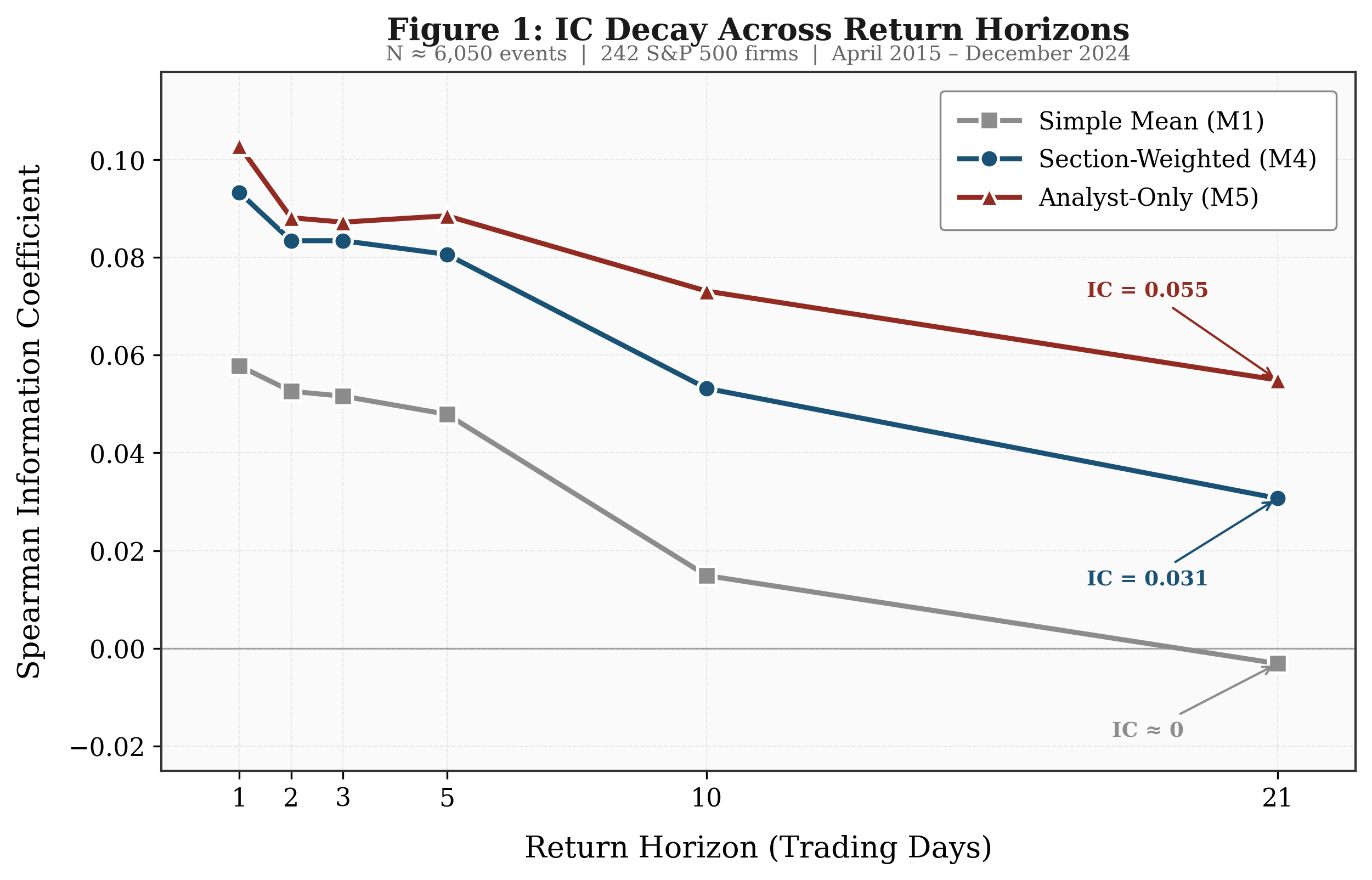}
\caption{Monthly Information Coefficients Over Time. Blue bars: in-sample (2015--2022); red bars: out-of-sample (2023+). Black line: six-month rolling average. In-sample mean IC = 0.099; OOS mean IC = 0.115. The signal is positive in 85\% of months and strengthens out-of-sample.}
\end{figure}

\subsection{Quintile Portfolio Analysis}

\begin{table}[!htbp]
\centering
\caption{Quintile Portfolio Returns --- Panel A: Full Sample ($N = 16{,}428$)}
\renewcommand{\arraystretch}{1.4}
\resizebox{\columnwidth}{!}{%
\begin{tabular}{|>{\centering\arraybackslash}m{1.8cm}|>{\centering\arraybackslash}m{0.8cm}|>{\centering\arraybackslash}m{0.7cm}|>{\centering\arraybackslash}m{0.7cm}|>{\centering\arraybackslash}m{0.7cm}|>{\centering\arraybackslash}m{0.8cm}|>{\centering\arraybackslash}m{0.9cm}|>{\centering\arraybackslash}m{0.7cm}|>{\centering\arraybackslash}m{0.7cm}|}
\hline
\rowcolor{headerblue}
\color{white}\textbf{Method} & \color{white}\textbf{Q1} & \color{white}\textbf{Q2} & \color{white}\textbf{Q3} & \color{white}\textbf{Q4} & \color{white}\textbf{Q5} & \color{white}\textbf{Q5-Q1} & \color{white}\textbf{$t$} & \color{white}\textbf{Mon} \\
\hline
\rowcolor{lightblue}
Simple (M1) & $-$0.67 & 0.24 & 0.34 & 0.38 & 0.93 & 1.60*** & 9.89 & Yes \\
\hline
Sect-wtd (M4) & $-$0.96 & $-$0.18 & 0.46 & 0.60 & 1.30 & 2.26*** & 13.33 & Yes \\
\hline
\rowcolor{lightblue}
Analyst (M5) & $-$1.22 & $-$0.10 & 0.19 & 0.77 & 1.63 & 2.85*** & 16.35 & Yes \\
\hline
\end{tabular}%
}
\end{table}

\begin{table}[!htbp]
\centering
\caption{Quintile Portfolio Returns --- Panel B: Out-of-Sample ($N = 5{,}131$)}
\renewcommand{\arraystretch}{1.4}
\resizebox{\columnwidth}{!}{%
\begin{tabular}{|>{\centering\arraybackslash}m{1.8cm}|>{\centering\arraybackslash}m{0.8cm}|>{\centering\arraybackslash}m{0.8cm}|>{\centering\arraybackslash}m{0.7cm}|>{\centering\arraybackslash}m{0.7cm}|>{\centering\arraybackslash}m{0.8cm}|>{\centering\arraybackslash}m{0.9cm}|>{\centering\arraybackslash}m{0.7cm}|>{\centering\arraybackslash}m{0.7cm}|}
\hline
\rowcolor{headerblue}
\color{white}\textbf{Method} & \color{white}\textbf{Q1} & \color{white}\textbf{Q2} & \color{white}\textbf{Q3} & \color{white}\textbf{Q4} & \color{white}\textbf{Q5} & \color{white}\textbf{Q5-Q1} & \color{white}\textbf{$t$} & \color{white}\textbf{Mon} \\
\hline
\rowcolor{lightblue}
Simple (M1) & $-$1.34 & $-$0.04 & 0.06 & 0.47 & 0.79 & 2.13*** & 6.28 & Yes \\
\hline
Sect-wtd (M4) & $-$1.77 & $-$0.38 & 0.31 & 0.40 & 1.38 & 3.14*** & 9.07 & Yes \\
\hline
\rowcolor{lightblue}
Analyst (M5) & $-$2.02 & $-$0.77 & $-$0.00 & 0.98 & 1.74 & 3.76*** & 10.58 & Yes \\
\hline
\end{tabular}%
}
\end{table}

\begin{table}[!htbp]
\centering
\caption{Panel C: Detailed Quintile Breakdown --- Section-Weighted (M4), Full Sample}
\renewcommand{\arraystretch}{1.35}
\resizebox{\columnwidth}{!}{%
\begin{tabular}{|>{\bfseries\centering\arraybackslash}m{2.0cm}|>{\centering\arraybackslash}m{0.8cm}|>{\centering\arraybackslash}m{1.2cm}|>{\centering\arraybackslash}m{1.2cm}|>{\centering\arraybackslash}m{1.2cm}|>{\centering\arraybackslash}m{1.0cm}|}
\hline
\rowcolor{headerblue}
\color{white}\textbf{Quintile} & \color{white}\textbf{$N$} & \color{white}\textbf{Mean sent.} & \color{white}\textbf{Mean ret.\ (\%)} & \color{white}\textbf{Std ret.\ (\%)} & \color{white}\textbf{$t$-stat} \\
\hline
\rowcolor{lightblue}
Q1 (most neg) & 3,286 & 0.0907 & $-$0.962 & 6.716 & $-$8.21 \\
\hline
Q2 & 3,285 & 0.1557 & $-$0.183 & 6.351 & $-$1.65 \\
\hline
\rowcolor{lightblue}
Q3 & 3,286 & 0.1941 & 0.458 & 6.916 & 3.79 \\
\hline
Q4 & 3,285 & 0.2310 & 0.603 & 6.853 & 5.04 \\
\hline
\rowcolor{lightblue}
Q5 (most pos) & 3,286 & 0.2906 & 1.301 & 7.042 & 10.59 \\
\hline
Q5 $-$ Q1 & & & 2.263*** & & 13.33 \\
\hline
\end{tabular}%
}
\tablenote{This table reports average 1-day post-earnings returns (\%) for quintile portfolios sorted by sentiment. Q1 contains the most negative sentiment calls; Q5 contains the most positive. Q5--Q1 is the long--short spread. Panel~A reports full-sample results. Panel~B reports out-of-sample results (January 2023 onward), using section weights frozen from the training period. Panel~C provides a detailed breakdown for M4, including the mean sentiment score, return standard deviation, and individual quintile $t$-statistics. Mono indicates monotonicity. The $t$-statistic for Q5--Q1 is from a two-sample $t$-test. *** $p < 0.01$.}
\end{table}

Table~8-10 shows the average one-day post-earnings returns for quintile portfolios organized by section-weighted emotion. Returns are precisely monotonic across quintiles, with Q1 (most negative sentiment) earning an average return of $-$0.96\% and Q5 (most positive) earning $+$1.30\%. The Q5--Q1 long--short spread is 2.26\% daily ($t = 13.33$). This monotonic pattern is economically significant. A quintile-sorted spread of more than 2\% in a single trading day and it supports Hypothesis~2.

In the out-of-sample time, the spread increases to 3.14\% ($t = 9.07$), demonstrating that the signal extends beyond the training data. Panel~C's detailed quintile breakdown reveals that both sides of the deal contribute: Q1 gets significantly negative returns ($t = -8.21$), while Q5 earns significantly positive returns.

\subsection{Controlling for Earnings Surprise}

\begin{table}[!htbp]
\centering
\caption{Controlling for Earnings Surprise: Panel A --- Information Coefficients}
\renewcommand{\arraystretch}{1.35}
\resizebox{\columnwidth}{!}{%
\begin{tabular}{|>{\centering\arraybackslash}m{2.5cm}|>{\centering\arraybackslash}m{1.2cm}|>{\centering\arraybackslash}m{1.2cm}|>{\centering\arraybackslash}m{1.5cm}|>{\centering\arraybackslash}m{1.0cm}|}
\hline
\rowcolor{headerblue}
\color{white}\textbf{Signal} & \color{white}\textbf{IC(1d)} & \color{white}\textbf{$t_{\text{NW}}$} & \color{white}\textbf{$p$-value} & \color{white}\textbf{$N$} \\
\hline
\rowcolor{lightblue}
SUE & 0.2458 & 17.58*** & 3.29e--69 & 16,109 \\
\hline
Sect-wtd (M4) & 0.1184 & 8.95*** & 3.55e--19 & 16,109 \\
\hline
\rowcolor{lightblue}
Analyst (M5) & 0.1409 & 11.94*** & 7.07e--33 & 15,954 \\
\hline
\end{tabular}%
}
\end{table}

\begin{table}[!htbp]
\centering
\caption{Controlling for Earnings Surprise: Panel B --- Fama--MacBeth Regressions}
\renewcommand{\arraystretch}{1.35}
\resizebox{\columnwidth}{!}{%
\begin{tabular}{|>{\centering\arraybackslash}m{2.4cm}|>{\centering\arraybackslash}m{1.4cm}|>{\centering\arraybackslash}m{1.4cm}|>{\centering\arraybackslash}m{1.4cm}|>{\centering\arraybackslash}m{1.4cm}|}
\hline
\rowcolor{headerblue}
\color{white}\textbf{Variable} & \color{white}\textbf{Model 1} & \color{white}\textbf{Model 2} & \color{white}\textbf{Model 3} & \color{white}\textbf{Model 4} \\
\hline
\rowcolor{lightblue}
Sect-wtd (M4) & 0.00679*** & --- & 0.00505*** & --- \\
& (7.16) & & (5.57) & \\
\hline
Analyst (M5) & --- & --- & --- & 0.00809*** \\
& & & & (6.94) \\
\hline
\rowcolor{lightblue}
SUE & --- & 0.01425*** & 0.01342*** & 0.01346*** \\
& & (14.17) & (13.43) & (13.17) \\
\hline
Months & 110 & 110 & 110 & 110 \\
\hline
\end{tabular}%
}
\tablenote{This table examines whether sentiment predicts post-earnings returns after controlling for SUE. Panel~A reports Spearman ICs. Panel~B reports Fama--MacBeth cross-sectional regression coefficients (standardized cross-sectionally), with Newey--West $t$-statistics in parentheses. Model~3 is the key specification: section-weighted sentiment remains highly significant ($t = 5.57$) after jointly controlling for SUE. The SUE coefficient declines only modestly from 0.01425 to 0.01342, confirming that sentiment and SUE capture largely orthogonal information. *** $p < 0.01$.}
\end{table}

A natural fear is that earnings call attitude only serves as a proxy for the hard earnings surprise: managers sound optimistic when they exceed expectations and depressed when they fall short. Table~11 handles this issue directly using Fama--MacBeth cross-sectional regressions. In the univariate model, section-weighted sentiment has a Fama--MacBeth coefficient of 0.00679 ($t = 7.16$). Table 11-12 shows when SUE is added as a joint regressor, the sentiment coefficient drops to 0.00505, although it remains highly significant ($t = 5.57$). The SUE coefficient keeps its predictive effectiveness after correcting for sentiment, dropping very slightly from 0.01425 (univariate) to 0.01342 (joint). These findings suggest that sentiment and SUE capture primarily orthogonal information, which supports Hypothesis~3.

\begin{table}[!htbp]
\centering
\caption{Double-Sort: Sentiment Quintiles Within SUE Terciles}
\renewcommand{\arraystretch}{1.35}
\resizebox{\columnwidth}{!}{%
\begin{tabular}{|>{\bfseries\centering\arraybackslash}m{2.0cm}|>{\centering\arraybackslash}m{0.8cm}|>{\centering\arraybackslash}m{1.3cm}|>{\centering\arraybackslash}m{1.3cm}|>{\centering\arraybackslash}m{1.3cm}|>{\centering\arraybackslash}m{0.8cm}|}
\hline
\rowcolor{headerblue}
\color{white}\textbf{SUE Tercile} & \color{white}\textbf{$N$} & \color{white}\textbf{Q1 (\%)} & \color{white}\textbf{Q5 (\%)} & \color{white}\textbf{Q5-Q1} & \color{white}\textbf{$t$} \\
\hline
\rowcolor{lightblue}
Low SUE & 5,370 & $-$2.767 & $-$0.539 & 2.227*** & 7.27 \\
\hline
Mid SUE & 5,369 & $-$0.291 & 1.323 & 1.613*** & 5.67 \\
\hline
\rowcolor{lightblue}
High SUE & 5,370 & 1.178 & 2.323 & 1.145*** & 4.06 \\
\hline
\end{tabular}%
}
\tablenote{This table reports a non-parametric test of whether sentiment predicts returns independently of earnings surprise. Observations are first sorted into terciles by SUE, then within each SUE tercile, stocks are sorted into quintiles by section-weighted sentiment (M4). The sentiment spread is statistically significant at the 1\% level in all three SUE terciles. The spread is largest for Low SUE firms (2.23\%), suggesting that sentiment is particularly informative when earnings disappoint---positive tone during a bad quarter signals management confidence that the market underweights.}
\end{table}

Table~13 shows non-parametric confirmation using double-sort analysis. Within each SUE tercile, the Q5--Q1 sentiment spread is positive and statistically significant at the 1\% level: 2.23\% for low-SUE enterprises, 1.61\% for mid-SUE firms, and 1.15\% for high-SUE firms. The highest spread occurs among firms with negative earnings surprises, implying that sentiment is especially useful when earnings disappoint. A positive tone during a difficult quarter conveys management confidence that the market is initially underweighted.

\begin{figure}[!htbp]
\centering
\includegraphics[width=0.9\columnwidth]{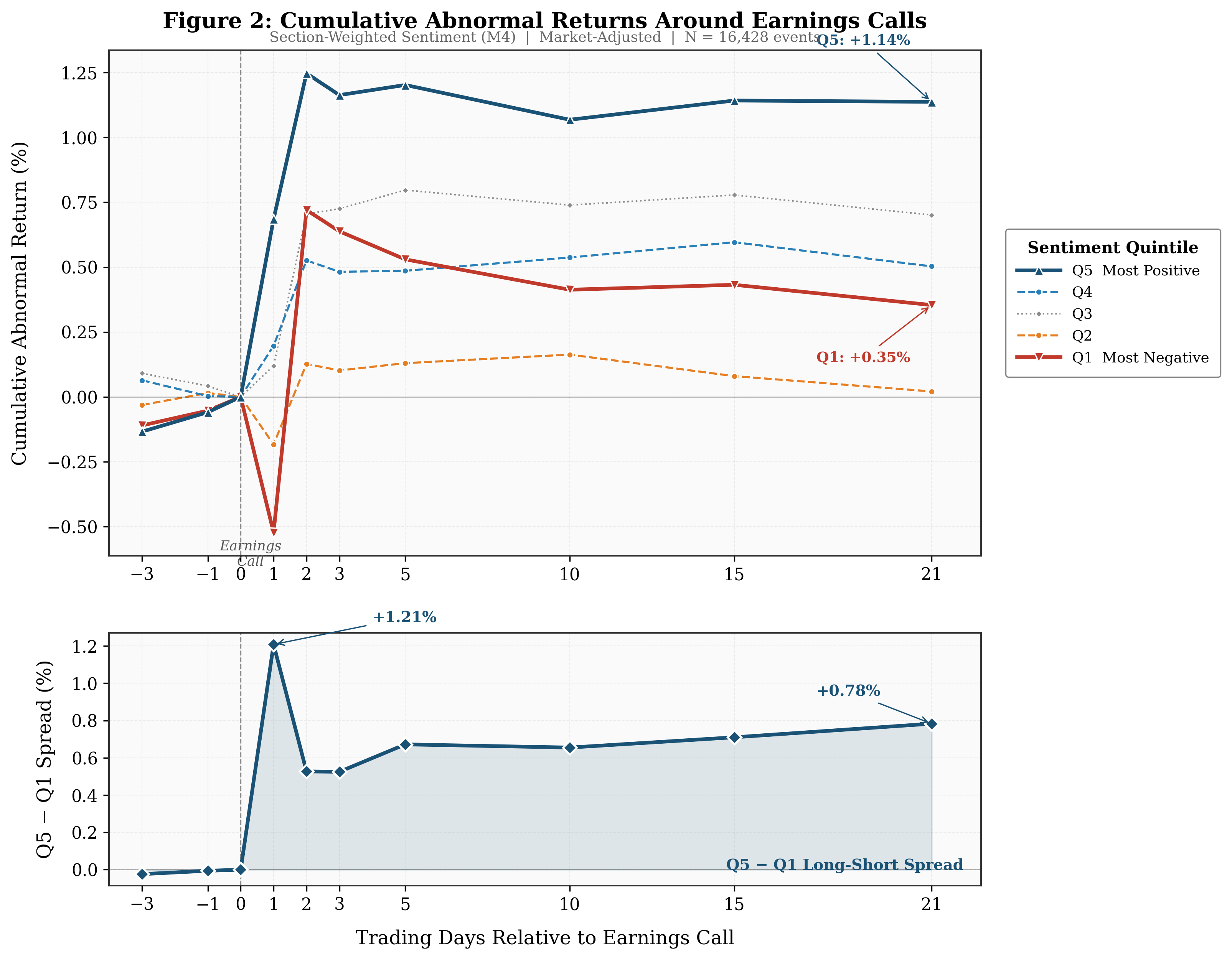}
\caption{Return Surface: Section-Weighted Sentiment $\times$ SUE. Mean 1-day post-earnings returns across 16,109 calls sorted into sentiment quintiles and SUE terciles. Surface spans $-2.63\%$ (Q1, Low SUE) to $+2.23\%$ (Q5, High SUE). Both dimensions contribute independently ($p < 0.01$).}
\end{figure}

\subsection{Fama--French Five-Factor Alpha}

\begin{table}[!htbp]
\centering
\caption{FF5 Alpha --- Panel A: Long--Short Portfolio Factor Loadings}
\renewcommand{\arraystretch}{1.35}
\resizebox{\columnwidth}{!}{%
\begin{tabular}{|>{\centering\arraybackslash}m{2.0cm}|>{\centering\arraybackslash}m{1.3cm}|>{\centering\arraybackslash}m{1.0cm}|>{\centering\arraybackslash}m{1.3cm}|>{\centering\arraybackslash}m{1.0cm}|}
\hline
\rowcolor{headerblue}
\color{white}\textbf{Factor} & \color{white}\textbf{M4 Coeff} & \color{white}\textbf{$t_{\text{NW}}$} & \color{white}\textbf{M5 Coeff} & \color{white}\textbf{$t_{\text{NW}}$} \\
\hline
\rowcolor{lightblue}
$\alpha$ (monthly) & 2.026\%*** & 6.49 & 2.542\%*** & 8.69 \\
\hline
$\alpha$ (annualized) & 24.31\% & & 30.51\% & \\
\hline
\rowcolor{lightblue}
Mkt--RF & 0.1223* & 1.74 & 0.1013 & 1.56 \\
\hline
SMB & $-$0.1816* & $-$1.79 & $-$0.0925 & $-$0.86 \\
\hline
\rowcolor{lightblue}
HML & 0.2002* & 1.86 & $-$0.0012 & $-$0.01 \\
\hline
RMW & $-$0.1077 & $-$0.72 & $-$0.1411 & $-$1.21 \\
\hline
\rowcolor{lightblue}
CMA & $-$0.1953 & $-$1.60 & 0.0858 & 0.59 \\
\hline
$R^2$ & 0.079 & & 0.034 & \\
\hline
\rowcolor{lightblue}
$N$ (months) & 100 & & 99 & \\
\hline
\end{tabular}%
}
\tablenote{The $R^2$ of 0.079 confirms the signal is largely orthogonal to known risk factors. Newey--West standard errors with $\lfloor 0.75 \cdot T^{1/3} \rfloor$ lags. *** $p < 0.01$; ** $p < 0.05$; * $p < 0.10$.}
\end{table}

Table~14 shows the findings of time-series regressions of sentiment-sorted portfolio returns on the Fama and French \cite{Fama2015} five factors. The long--short portfolio based on section-weighted sentiment generates a monthly alpha of 2.03\% ($t = 6.49$), which translates to an annualized alpha of 24.3\%. The five-factor regression has an $R^2$ of only 0.079, indicating that known risk variables explain little of the long--short return variation. Factor loadings are tiny and largely inconsequential, indicating that the emotion signal is not a proxy for any of the usual risk factors.

\begin{table}[!htbp]
\centering
\caption{FF5 Alpha --- Panel B: Alpha by Quintile, Section-Weighted (M4)}
\renewcommand{\arraystretch}{1.35}
\resizebox{\columnwidth}{!}{%
\begin{tabular}{|>{\bfseries\centering\arraybackslash}m{1.8cm}|>{\centering\arraybackslash}m{1.3cm}|>{\centering\arraybackslash}m{1.3cm}|>{\centering\arraybackslash}m{1.0cm}|>{\centering\arraybackslash}m{1.3cm}|}
\hline
\rowcolor{headerblue}
\color{white}\textbf{Quintile} & \color{white}\textbf{$\alpha$ (mo.)} & \color{white}\textbf{$\alpha$ (ann.)} & \color{white}\textbf{$t_{\text{NW}}$} & \color{white}\textbf{$p$-value} \\
\hline
\rowcolor{lightblue}
Q1 (most neg) & $-$0.939\%*** & $-$11.27\% & $-$3.34 & 8.41e--04 \\
\hline
Q2 & 0.040\% & 0.48\% & 0.21 & 8.31e--01 \\
\hline
\rowcolor{lightblue}
Q3 & 0.106\% & 1.27\% & 0.49 & 6.25e--01 \\
\hline
Q4 & 0.479\%*** & 5.74\% & 3.36 & 7.94e--04 \\
\hline
\rowcolor{lightblue}
Q5 (most pos) & 1.086\%*** & 13.04\% & 7.76 & 8.75e--15 \\
\hline
Q5 $-$ Q1 & 2.026\%*** & 24.31\% & 6.49 & 8.36e--11 \\
\hline
\end{tabular}%
}
\end{table}

Table 15 decomposes alpha into distinct quintiles. The pattern is strikingly monotonic: Q1 gets a monthly alpha of $-$0.94\% ($t = -3.34$), Q2 through Q3 earn near-zero alpha, and Q5 earns $+$1.09\% ($t = 7.76$). The fact that both the long and short sides contribute to the spread contradicts risk-based theories and supports the view that the mood signal captures mispricing caused by the sluggish integration of soft information.

\begin{table}[!htbp]
\centering
\caption{Sub-Period and Out-of-Sample FF5 Alpha}
\renewcommand{\arraystretch}{1.35}
\resizebox{\columnwidth}{!}{%
\begin{tabular}{|>{\centering\arraybackslash}m{2.2cm}|>{\centering\arraybackslash}m{1.2cm}|>{\centering\arraybackslash}m{1.0cm}|>{\centering\arraybackslash}m{0.8cm}|>{\centering\arraybackslash}m{1.2cm}|>{\centering\arraybackslash}m{0.8cm}|}
\hline
\rowcolor{headerblue}
\color{white}\textbf{Period} & \color{white}\textbf{$\alpha$ (mo.)} & \color{white}\textbf{$\alpha$ (ann.)} & \color{white}\textbf{$t_{\text{NW}}$} & \color{white}\textbf{$p$-value} & \color{white}\textbf{Mo.} \\
\hline
\multicolumn{6}{|l|}{\cellcolor{row1}\textit{Sub-period analysis}} \\
\hline
\rowcolor{lightblue}
Pre-COVID (15--19) & 2.227\%*** & 26.73\% & 11.70 & 1.33e--31 & 39 \\
\hline
COVID (20--21) & $-$0.505\% & $-$6.06\% & $-$0.68 & 4.99e--01 & 21 \\
\hline
\rowcolor{lightblue}
Post-COVID (22+) & 2.865\%*** & 34.38\% & 10.35 & 4.29e--25 & 40 \\
\hline
\multicolumn{6}{|l|}{\cellcolor{row1}\textit{Out-of-sample validation}} \\
\hline
\rowcolor{lightblue}
Train (pre-23) & 1.560\%*** & 18.72\% & 4.48 & 7.55e--06 & 71 \\
\hline
Test (2023+) & 2.770\%*** & 33.23\% & 8.56 & 1.16e--17 & 29 \\
\hline
\end{tabular}%
}
\tablenote{This table reports the FF5 alpha of the section-weighted (M4) long--short portfolio across sub-periods and out-of-sample using weights frozen from the training period. Alpha is significant at the 1\% level in all periods except 2020--2021. The test-period alpha (2.77\%/month, $t = 8.56$) exceeds the training-period alpha (1.56\%/month), consistent with out-of-sample signal strengthening. Newey--West standard errors with $\lfloor 0.75 \cdot T^{1/3} \rfloor$ lags.}
\end{table}

Table~16 investigates alpha stability across subperiods. The section-weighted long--short portfolio provides significant alpha both before and after COVID (2.23\%/month, $t = 11.70$ and 2.87\%/month, $t = 10.35$). The only time with low alpha is 2020--2021, when enormous macroeconomic actions dominated individual stock price movements. In the out-of-sample test period (2023 onward), alpha is 2.77\% per month ($t = 8.56$), which exceeds the training-period alpha of 1.56\%, consistent with the IC evidence of out-of-sample progress.

\subsection{FinBERT versus Loughran--McDonald: Horse Race}

\begin{table}[!htbp]
\centering
\caption{FinBERT vs.\ LM: Panel A --- Full-Sample IC}
\renewcommand{\arraystretch}{1.4}
\resizebox{\columnwidth}{!}{%
\begin{tabular}{|>{\centering\arraybackslash}m{1.8cm}|>{\centering\arraybackslash}m{0.8cm}|>{\centering\arraybackslash}m{0.9cm}|>{\centering\arraybackslash}m{0.8cm}|>{\centering\arraybackslash}m{0.9cm}|>{\centering\arraybackslash}m{1.0cm}|}
\hline
\rowcolor{headerblue}
\color{white}\textbf{Method} & \color{white}\textbf{FB IC} & \color{white}\textbf{FB $t$} & \color{white}\textbf{LM IC} & \color{white}\textbf{LM $t$} & \color{white}\textbf{FB/LM} \\
\hline
\rowcolor{lightblue}
Simple mean & 0.0813 & 10.46*** & 0.0691 & 8.88*** & 1.2$\times$ \\
\hline
Conf-wtd & 0.0986 & 12.70*** & 0.0930 & 11.98*** & 1.1$\times$ \\
\hline
\rowcolor{lightblue}
Extreme & 0.0749 & 9.63*** & 0.0716 & 9.20*** & 1.0$\times$ \\
\hline
Section-wtd & 0.1188 & 15.33*** & 0.0745 & 9.58*** & 1.6$\times$ \\
\hline
\rowcolor{lightblue}
Analyst-only & 0.1405 & 18.10*** & 0.0619 & 7.91*** & 2.3$\times$ \\
\hline
\end{tabular}%
}
\tablenote{Full-sample Spearman ICs against 1-day post-earnings returns ($N = 16{,}428$). FinBERT dominates LM across all methods; the advantage is largest for section-weighted (1.6$\times$) and analyst-only (2.3$\times$) aggregations. *** $p < 0.01$.}
\end{table}

\begin{table}[!htbp]
\centering
\caption{FinBERT vs.\ LM: Panel B --- Out-of-Sample IC}
\renewcommand{\arraystretch}{1.4}
\resizebox{\columnwidth}{!}{%
\begin{tabular}{|>{\centering\arraybackslash}m{1.8cm}|>{\centering\arraybackslash}m{1.0cm}|>{\centering\arraybackslash}m{1.0cm}|>{\centering\arraybackslash}m{1.0cm}|>{\centering\arraybackslash}m{1.0cm}|}
\hline
\rowcolor{headerblue}
\color{white}\textbf{Method} & \color{white}\textbf{FB Train} & \color{white}\textbf{FB Test} & \color{white}\textbf{LM Train} & \color{white}\textbf{LM Test} \\
\hline
\rowcolor{lightblue}
Simple mean & 0.0806 & 0.0966 & 0.0625 & 0.0876 \\
\hline
Conf-wtd & 0.1005 & 0.1075 & 0.0963 & 0.0944 \\
\hline
\rowcolor{lightblue}
Extreme & 0.0736 & 0.0914 & 0.0765 & 0.0763 \\
\hline
Section-wtd & 0.1141 & 0.1442 & 0.0586 & 0.1077 \\
\hline
\rowcolor{lightblue}
Analyst-only & 0.1273 & 0.1707 & 0.0452 & 0.0936 \\
\hline
\end{tabular}%
}
\tablenote{FinBERT's OOS advantage widens for section-weighted and analyst-only methods. Train: $N = 11{,}297$; Test: $N = 5{,}131$.}
\end{table}

\begin{table}[!htbp]
\centering
\caption{FinBERT vs.\ LM: Panel C --- Quintile Long--Short Spreads}
\renewcommand{\arraystretch}{1.4}
\begin{tabular}{|>{\centering\arraybackslash}m{2.5cm}|>{\centering\arraybackslash}m{2.0cm}|>{\centering\arraybackslash}m{2.0cm}|}
\hline
\rowcolor{headerblue}
\color{white}\textbf{Method} & \color{white}\textbf{FinBERT Q5-Q1} & \color{white}\textbf{LM Q5-Q1} \\
\hline
\rowcolor{lightblue}
Simple mean & 1.604\% & 1.251\% \\
\hline
Conf-wtd & 1.844\% & 1.678\% \\
\hline
\rowcolor{lightblue}
Extreme & 1.469\% & 1.060\% \\
\hline
Section-wtd & 2.263\% & 1.355\% \\
\hline
\rowcolor{lightblue}
Analyst-only & 2.847\% & 1.029\% \\
\hline
\end{tabular}
\end{table}

Table~17 shows a head-to-head comparison of FinBERT and Loughran--McDonald (LM) prediction power for all five aggregations. FinBERT outperforms LM across all methods, with the greatest advantage for section-weighted ($\mathrm{IC}$: 0.119 vs.\ 0.075, a 1.6$\times$ ratio) and analyst-only ($\mathrm{IC}$: 0.141 vs.\ 0.062, a 2.3$\times$ ratio) aggregations. FinBERT's contextual awareness is most effective when applied to the language of sophisticated market players (analysts) who use hedging, conditional wording, and implicit comparison. Tables~ 18 and 19 confirm that FinBERT's advantage widens out-of-sample and translates into larger quintile spreads. The advantage is weakest for confidence-weighted aggregate (1.1$\times$).

\begin{table}[!htbp]
\centering
\caption{Fama--MacBeth Horse Race: Does FinBERT Subsume LM?}
\renewcommand{\arraystretch}{1.35}
\resizebox{\columnwidth}{!}{%
\begin{tabular}{|>{\centering\arraybackslash}m{2.8cm}|>{\centering\arraybackslash}m{1.4cm}|>{\centering\arraybackslash}m{1.4cm}|>{\centering\arraybackslash}m{1.4cm}|}
\hline
\rowcolor{headerblue}
\color{white}\textbf{Variable} & \color{white}\textbf{Model 1 (FB)} & \color{white}\textbf{Model 2 (LM)} & \color{white}\textbf{Model 3 (Joint)} \\
\hline
\multicolumn{4}{|l|}{\cellcolor{row1}\textit{Panel A: Section-Weighted Sentiment}} \\
\hline
\rowcolor{lightblue}
FinBERT sect-wtd & 0.00692*** & --- & 0.00660*** \\
& (7.14) & & (5.90) \\
\hline
LM sect-wtd & --- & 0.00306*** & 0.00073 \\
& & (4.09) & (0.86) \\
\hline
\multicolumn{4}{|l|}{\cellcolor{row1}\textit{Panel B: Analyst-Only Sentiment}} \\
\hline
\rowcolor{lightblue}
FinBERT analyst & 0.00956*** & --- & 0.00986*** \\
& (8.18) & & (7.65) \\
\hline
LM analyst & --- & 0.00262*** & $-$0.00109 \\
& & (3.57) & ($-$1.37) \\
\hline
\rowcolor{lightblue}
Months & 112 & 112 & 112 \\
\hline
\end{tabular}%
}
\tablenote{This table reports Fama--MacBeth cross-sectional regressions testing whether FinBERT sentiment subsumes the LM dictionary. Newey--West $t$-statistics in parentheses. Models 3 and 6 are the key subsumption tests. FinBERT remains highly significant ($t = 5.90$ and $t = 7.65$) while LM becomes insignificant ($t = 0.86$ and $t = -1.37$), demonstrating that FinBERT \textit{completely subsumes} the LM dictionary. The FinBERT coefficient is virtually unchanged between univariate and joint specifications ($0.00692 \rightarrow 0.00660$ in Panel~A; $0.00956 \rightarrow 0.00986$ in Panel~B), confirming LM adds no incremental information. *** $p < 0.01$.}
\end{table}

Table~20 shows the definitive subsumption test. In the Fama--MacBeth horse race with section-weighted aggregation, FinBERT alone gives a coefficient of 0.00692 ($t = 7.14$), while LM alone produces 0.00306 ($t = 4.09$). When both are considered together, FinBERT maintains its significance (0.00660, $t = 5.90$), whereas LM becomes statistically indistinguishable from zero (0.00073, $t = 0.86$). The results are even more dramatic for analyst-only aggregation: in the joint specification, FinBERT's coefficient actually grows somewhat (0.00986, $t = 7.65$), whereas LM's becomes negative and negligible ($-$0.00109, $t = -1.37$). This comprehensive subsumption result shows that FinBERT catches all the dictionary does, plus significant more information gleaned from comprehending the linguistic context in which financial terms appear. This finding supports Hypothesis~4.

\FloatBarrier
\section{Additional Tests and Robustness}

\subsection{Return Horizon Decay}

We look at how the Information Coefficient of the section-weighted sentiment signal decays over return horizons ranging from 1 to 21 trading days. The IC is maximum for the one-day horizon (0.119) and decreases almost exponentially, reaching 0.099 after 5 days, 0.065 after 10 days, and 0.032 after 21 days. The signal has a half-life of about 6--7 trading days, which supports the theory that earnings call emotion comprises short-lived information that is gradually integrated into prices throughout the first one to two weeks after the call. This decay pattern is significant for practitioners since the signal is most beneficial when executed shortly after the call and fades over a monthly period.

\subsection{Cumulative Abnormal Returns}

We create cumulative abnormal return (CAR) charts by categorizing companies into sentiment quintiles during each earnings call and tracking abnormal returns (compared to the S\&P 500) for 30 trading days after the call. The CAR chart shows the classic fan-shaped divergence between Q1 and Q5. Over 30 days, the most positive sentiment quintile collects around $+$1.5\% in anomalous returns, while the most negative quintile accumulates approximately $-$1.5\%, resulting in a total variance of around 3.0\%. The divergence begins abruptly on day 0 (the announcement day) and steadily widens over the next two weeks, consistent with delayed information assimilation and the PEAD literature.

\subsection{Industry Controls}

To ensure that our findings are not influenced by industry-level sentiment clustering for example, if technology businesses consistently have more positive earnings calls and greater returns. We incorporate GICS sector fixed effects into the Fama--MacBeth regressions. The section-weighted sentiment coefficient remains significant following this adjustment, indicating that the signal acts within industries rather than across them.

\subsection{Machine Learning Validation: XGBoost with SHAP}

As an alternative validation of the IC-derived section weights, we train an XGBoost gradient-boosted tree model\cite{Chen2016} to predict one-day post-earnings returns using the four speaker-category sentiment scores as input features. The model is trained on the same pre-2023 sample and evaluated out-of-sample with frozen parameters. We then apply SHAP\cite{Lundberg2017} (SHapley Additive exPlanations) to decompose the model's predictions into feature-level contributions on the held-out test set.

\begin{table}[!htbp]
\centering
\caption{XGBoost--SHAP Feature Importance and IC Validation}
\renewcommand{\arraystretch}{1.35}
\resizebox{\columnwidth}{!}{%
\begin{tabular}{|>{\bfseries\centering\arraybackslash}m{2.8cm}|>{\centering\arraybackslash}m{1.5cm}|>{\centering\arraybackslash}m{1.3cm}|>{\centering\arraybackslash}m{1.3cm}|}
\hline
\rowcolor{headerblue}
\color{white}\textbf{Speaker Feature} & \color{white}\textbf{Mean $|$SHAP$|$} & \color{white}\textbf{SHAP \%} & \color{white}\textbf{IC Wt \%} \\
\hline
\rowcolor{lightblue}
Analyst & 0.00489 & 59.6\% & 48.8\% \\
\hline
CFO / Finance & 0.00239 & 29.1\% & 29.5\% \\
\hline
\rowcolor{lightblue}
Top Executive & 0.00070 & 8.5\% & 15.9\% \\
\hline
Other Mgmt & 0.00023 & 2.8\% & 5.8\% \\
\hline
\multicolumn{4}{|l|}{\cellcolor{row1}\textit{Model Information Coefficients}} \\
\hline
\rowcolor{lightblue}
In-sample IC & \multicolumn{3}{c|}{0.1723\enspace(std $= 0.0613$)} \\
\hline
Out-of-sample IC & \multicolumn{3}{c|}{0.1711\enspace(std $= 0.0686$)} \\
\hline
\end{tabular}%
}
\tablenote{This table reports SHAP feature importance from an XGBoost model trained on the four speaker-category FinBERT sentiment scores to predict 1-day post-earnings returns. The model is trained on the pre-2023 sample and evaluated out-of-sample. Mean $|$SHAP$|$ is the average absolute SHAP value across test-set observations. SHAP \% is each feature's share of total importance. IC Wt \% reproduces the Spearman IC-derived weights from Table~5 for comparison. The SHAP ranking confirms the IC-derived hierarchy. The near-zero in-sample to out-of-sample IC decay (0.172 $\rightarrow$ 0.171) confirms the stability of the speaker importance structure.}
\end{table}

The SHAP-derived importance ranking closely mirrors the IC-derived weight hierarchy: analyst sentiment dominates with 59.6\% of total SHAP importance (compared to 48.8\% IC weight), followed by CFO sentiment at 29.1\% (vs.\ 29.5\% IC weight). The XGBoost model achieves an in-sample Spearman IC of 0.172 (std $= 0.061$) and an out-of-sample IC of 0.171 (std $= 0.069$), representing a meaningful improvement over the linear section-weighted approach (OOS IC $= 0.144$). Crucially, the near-zero decay between in-sample and out-of-sample IC (0.172 vs.\ 0.171) further confirms that the speaker-level sentiment hierarchy is a stable, structural feature of the earnings call information environment rather than a statistical artifact. The convergence between the model-free IC weighting approach and the nonlinear machine learning model provides strong corroborating evidence that analyst sentiment is the dominant return-predictive component of earnings calls, and that the relative ranking of speaker categories is robust to the choice of estimation methodology.

\subsection{Example Sentences: FinBERT versus Loughran--McDonald}

Table~22 illustrates the divergence between FinBERT and LM classification. Panel~A shows cases where FinBERT detects clear sentiment but LM finds no dictionary words for example, ``15\% YoY growth'' scores $+$0.95 in FinBERT but 0.00 in LM because ``growth'' is absent from the LM lexicon. Panel~B highlights disagreements: LM flags ``strong'' as positive in ``down 24\% from strong performance,'' while FinBERT correctly reads the sentence as negative. Panel~C confirms both methods agree on unambiguous cases.

\begin{table}[H]
\centering
\caption{Example Sentences: FinBERT vs.\ LM Divergence}
\renewcommand{\arraystretch}{1.25}
\resizebox{\columnwidth}{!}{%
\begin{tabular}{|>{\arraybackslash}m{4.2cm}|>{\centering\arraybackslash}m{0.8cm}|>{\centering\arraybackslash}m{0.7cm}|>{\centering\arraybackslash}m{1.5cm}|}
\hline
\rowcolor{headerblue}
\color{white}\textbf{Sentence (abridged)} & \color{white}\textbf{FB} & \color{white}\textbf{LM} & \color{white}\textbf{Category} \\
\hline
\multicolumn{4}{|l|}{\cellcolor{row1}\textit{Panel A: FinBERT detects, LM sees nothing}} \\
\hline
\rowcolor{lightblue}
``This 15\% YoY growth for JAKAFA net product sales reflects higher patient demand across all indications.'' & $+$0.95 & 0.00 & FB$+$ / LM$\,\emptyset$ \\
\hline
``Revenues were \$545M, down 5\% vs.\ prior year, while adjusted revenues fell 4\%.'' & $-$0.97 & 0.00 & FB$-$ / LM$\,\emptyset$ \\
\hline
\multicolumn{4}{|l|}{\cellcolor{row1}\textit{Panel B: FinBERT and LM disagree}} \\
\hline
\rowcolor{lightblue}
``Excluding negative FX, revenue grew 4\%\ldots\ while EPS increased 14\%.'' & $+$0.95 & $-$0.04 & FB$+$ / LM$-$ \\
\hline
``FICC revenues \$2B, down 24\% from strong performance last year.'' & $-$0.97 & $+$0.04 & FB$-$ / LM$+$ \\
\hline
\multicolumn{4}{|l|}{\cellcolor{row1}\textit{Panel C: Both agree (validation)}} \\
\hline
\rowcolor{lightblue}
``Strong global growth improved end-user demand across all regions.'' & $+$0.95 & $+$0.18 & Both $+$ \\
\hline
``Total revenue down 4\%, including 3\% organic decline and 1 ppt unfavorable FX.'' & $-$0.97 & $-$0.12 & Both $-$ \\
\hline
\end{tabular}%
}
\tablenote{FinBERT scores: $[-1, +1]$. LM: $(N_{\text{pos}} - N_{\text{neg}})/N_{\text{words}}$. Panel~A: FinBERT detects sentiment where LM finds no dictionary words. Panel~B: disagreements from context-blind word matching. Panel~C: agreement on unambiguous cases.}
\end{table}

\FloatBarrier
\section{Conclusion}

This paper demonstrates that the identity of the speaker in an earnings conference call matters as much as what they say for predicting post-announcement stock returns. By applying FinBERT to 6.5 million sentences from S\&P 500 earnings call transcripts and decomposing sentiment by speaker role, we show that an empirically derived section-weighting scheme, which assigns 49\% weight to analyst sentiment, 30\% to CFOs, 16\% to executives, and 5\% to other speakers, substantially outperforms na\"ive equal-weighted aggregation. The section-weighted sentiment signal achieves an out-of-sample Information Coefficient of 0.142, generates monthly Fama--French five-factor alpha of 2.03\% for the long--short portfolio, and remains significant after controlling for standardized unexpected earnings.

Our most important methodological finding is the complete subsumption of the Loughran--McDonald dictionary by FinBERT in a controlled horse-race regression. When both measures are included in the same Fama--MacBeth specification, FinBERT retains full statistical significance while the dictionary-based measure becomes indistinguishable from zero. This result has broad implications for the textual analysis literature: it suggests that context-aware deep-learning models capture everything that bag-of-words methods capture, plus substantial additional information arising from an understanding of negation, hedging, and comparative framing in financial language.

Several implications follow from our findings. For practitioners, our results identify a specific, implementable alpha source: buying stocks with positive earnings call sentiment and selling those with negative sentiment, with weights tilted toward analyst commentary. For academics, our section-weighted IC framework provides a principled methodology for constructing sentiment signals from any multi-speaker corporate communication. For regulators and market microstructure researchers, the finding that markets are slow to incorporate the nuanced information embedded in earnings call conversations, particularly the divergence between analyst and management tone, contributes to the growing understanding of how soft information is priced in equity markets.

Our analysis is subject to certain limitations. The sample is restricted to S\&P 500 firms, which represent the large-capitalization segment of the market; it remains an open question whether section-weighted sentiment is equally informative for small-cap firms where information asymmetry may be more acute. We also note that implementation of the long--short strategy would face transaction costs and capacity constraints that we do not model. Finally, while our out-of-sample results are encouraging, the test period (2023--2025) is relatively short, and longer-term stability of the signal warrants continued monitoring.

\end{document}